\documentclass[aps,prl,reprint,showpacs]{revtex4-1}
\usepackage[dvipdfmx]{graphicx}
\usepackage{bm}
\def\v#1{\mathbf{#1}} 
\usepackage{amsmath,amssymb}
\usepackage{color}
\usepackage{latexsym} 
\usepackage{bm}
\def\vb#1{{\bm#1}}
\def\v#1{\mathbf{#1}}			

\def\vr{\v{r}} 					
\def\vq{\v{q}} 					
\def\vk{\v{k}} 					

\def\vS{\v{v}_{\vb{\sigma}}}

\def\vS{\v{S}}

\def\vsigma{\vb{\sigma}}

\def\del{\partial} 

\def\la{\langle}
\def\ra{\rangle}

\def\kB{k_{\rm B}} 
\def\dm{\delta \mu_{\rm S}} 
\def\Jsd{J_{\rm sd}} 
\def\Nint{N_{\rm int}} 
\def\Jint{J^2_{\rm int}} 
\def\RE{{\rm Re}} 
\def\IM{{\rm Im}} 
\def\ikqom{\int_{\vq\vk\omega}} 
\def\Gr{G^R_{\vk\omega}}
\def\Ga{G^A_{\vk\omega}}
\def\Gl{G^<_{\vk\omega}}

\def\fFM{f^{\rm F}_{\omega}}

\def\xr{\chi^R_{\vq\omega}}

\def\xl{\chi^<_{\vq\omega}}

\def\fNM{f^{\rm P}_{\omega}}

\begin{document}

\title{
Theory of spin Peltier effect
}

\author{Y. Ohnuma$^{1}$,
M. Matsuo$^{1,2}$ and S. Maekawa$^{1}$}
\affiliation{%
${^1}$Advanced Science Research Center, Japan Atomic Energy Agency, Tokai 319-1195, Japan. \\
${^2}$Advanced Institute for Materials Research, Tohoku University, Sendai, 980-8577, Japan.
}%


\begin{abstract} 
A microscopic theory of the spin Peltier effect in a bilayer structure comprising a paramagnetic metal (PM) and a ferromagnetic insulator (FI) based on the nonequilibrium Green's function method is presented. Spin current and heat current driven by temperature gradient and spin accumulation are formulated as functions of spin susceptibilities in the PM and the FI, and are summarized by Onsager's reciprocal relations. 
By using the current formulae, we estimate heat generation and absorption at the interface driven by the heat-current injection mediated by spins from PM into FI. 
\end{abstract} 
\maketitle 

\paragraph{Introduction.---}
In the field of spintronics, 
inter-conversion between heat and spin current has attracted considerable attention and has been studied actively since the discovery of the spin Seebeck effect~\cite{Uchida08,Jaworski10,Uchida10}.
The spin Seebeck effect refers to the spin-current generation from heat in magnetic materials~\cite{Xiao10,Adachi11}. 
The spin Seebeck effect has been observed in a variety of materials ranging from magnetic metals and semiconductors to insulators~\cite{Uchida08,Jaworski10,Uchida10}.
Recently, the reciprocal phenomenon of the spin Seebeck effect, 
heat generation from spin current, was reported experimentally~\cite{Flipse14,Daimon16}. 
While the spin Peltier effect has been studied using a phenomenological model~\cite{Gravier06,Hatami09,Kovalev09,Bauer10,Basso16}, its microscopic theory is missing. 

\begin{figure}[!hbtp]
	\begin{center}
	\includegraphics[scale=0.4]{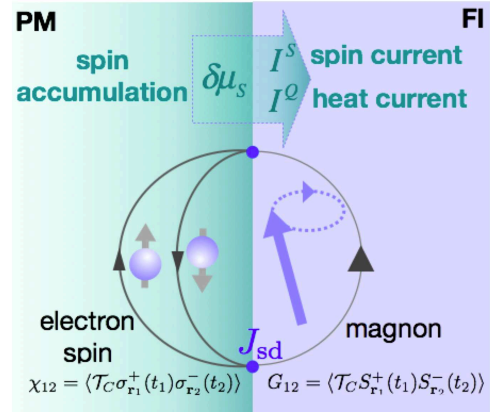}
	\caption{(Color online) Schematic view of the spin Peltier effect. We consider spin transport in a bilayer structure consisting of a paramagnetic metal (PM) and a ferromagnetic insulator (FI), where the electron spins in PI are coupled with the localized moments in FI via the exchange interaction $\Jsd$. The spin accumulation at the interface ($\delta \mu_s$) is found to be a driving force of spin and heat current ($I^S$ and $I^Q$) by using the nonequilibrium Green's functions for electron spin $\chi$ and magnon $G$, where $\mathcal{T}_C$ denotes the time ordering on the Keldysh contour. 
    }\label{fig1_model}
	\end{center}
\end{figure}

In this study, we formulate a microscopic theory of the spin Peltier effect in paramagnetic metal (PM)/ferromagnetic insulator (FI) junction systems by using the nonequilibrium Green's function method. 
To reveal the microscopic mechanism of spin and heat transfer, 
we perform investigations using the setup shown in Fig.~\ref{fig1_model}, where electron spins in PM, $\sigma$, are coupled with localized spins in FI, $S$, via the exchange interaction, $J_{sd}$. 
Let us consider spin accumulation at the interface, $\delta \mu_S$, generated by the spin Hall effect~\cite{Sinova16} in PM. Owing to the exchange interaction, this spin accumulation excites the localized spins in FI, and then, 
magnon flows are induced, accompanying both the spin and the heat. 

\paragraph{Spin-current generation at magnetic interface.---}
Let us briefly review spin-current generation in PM/FI by using the nonequilibrium Green's function. 
The magnetic interface is modeled using the s$-$d exchange interaction:
\begin{eqnarray}
	H_{\rm sd}
	&=& \Jsd \!\! \sum_{i\in {\rm int}} \vsigma_i\cdot\vS_i,
\label{Eq:model-sd}
\end{eqnarray}
where $\Jsd$, $\vsigma_i$ and $\vS_i$ represent the coupling constant of the exchange interaction, Pauli matrices and localized spin of FI, respectively, and $\sum_{i\in {\rm int}}$ denotes the summation on the lattice sites at the interface. 

The spin current $I^{\rm S}$ is defined by the time derivative of the z-component of the conduction electron spin in PM, that is, $I^{\rm S}\equiv\sum_{i\in P}\la \del_t \sigma^z_i \ra$, where $\la \cdots \ra$ denotes the statistical average~\cite{Konig03}. The Heisenberg equation of motion for $\sigma^z_i$ gives $I^{\rm S}=2\Jsd\hbar^{-1}\RE\sum_{i\in {\rm int}}\la\sigma^+_i(t) S^-_i(t)\ra$~\cite{Adachi11}, where $\sigma^{\pm}_i=\sigma^x_i \pm i\sigma^y_i$ and $S^{\pm}_i=S^x_i \pm i S^y_i$. 
After the perturbative calculation~\cite{Meir92,Haug-text} of $\la\sigma^+_i(t)S^-_i(t)\ra$ up to the second order of $\Jsd$, the spin current is given by
\begin{eqnarray}
	I^{\rm S}
	&=& 2\Jint\RE \!\! \ikqom \!\!\!\!\! (\chi^R_{\vq\vr,\omega t}G^<_{\vk\vr',\omega t}+\chi^<_{\vq\vr,\omega t}G^A_{\vk\vr',\omega t}),
\label{Eq:Spin-current-01}
\end{eqnarray}
where, we have introduced the shorthand notation $\ikqom=\int d^3\vk d^3\vq \int^{\infty}_{-\infty}\frac{d\omega}{2\pi}$. $\Jint$ is given by $\Jint=(\Jsd/\hbar)^2 \Nint$, with $\Nint$ being the number of sites at the interface. 
In Eq.~(\ref{Eq:Spin-current-01}), $\chi^{R(<)}_{\vq\vr,\omega t}$ is the retarded (lesser) component of the transverse spin susceptibility in PM given by $\chi^{R(<)}_{\vq\vr,\omega t}=\int_{\delta\vr\delta t}\exp[-i\vq\cdot\delta\vr+i\omega\delta t]\chi^{R(<)}(\vr+\delta\vr/2,t+\delta t;\vr-\delta\vr/2,t-\delta t)$, where $\chi^R(\vr+\delta\vr/2,t+\delta t;\vr-\delta\vr/2,t-\delta t)$ and $\chi^<(\vr+\delta\vr/2,t+\delta t;\vr-\delta\vr/2,t-\delta t)$ are defined as $\chi^R(\vr+\delta\vr/2,t+\delta t;\vr-\delta\vr/2,t-\delta t)\equiv-i\theta(t_1-t_2)\la [\sigma^+_{\vr_1}(t_1),\sigma^-_{\vr_2}(t_2)]\ra$ and 
$\chi^<(\vr+\delta\vr/2,t+\delta t;\vr-\delta\vr/2,t-\delta t)\equiv-i\la \sigma^+_{\vr_1}(t_1)\sigma^-_{\vr_2}(t_2)\ra$, respectively, with $\vr\equiv(\vr_1+\vr_2)/2$, $\delta\vr=\vr_1-\vr_2$, $t\equiv(t_1+t_2)/2$ and $\delta t=t_1-t_2$. 
$G^{A(<)}_{\vk\vr',\omega t}$ is the advanced (lesser) component of the transverse spin susceptibility in FI, and it is given by $G^{A(<)}_{\vk\vr',\omega t}=\int_{\delta\vr\delta t}\exp[-i\vk\cdot\delta\vr+i\omega\delta t]G^{A(<)}(\vr'+\delta\vr/2,t+\delta t;\vr'-\delta\vr/2,t-\delta t)$, where $G^A(\vr'+\delta\vr/2,t+\delta t;\vr'-\delta\vr/2,t-\delta t)$ and $G^<(\vr'+\delta\vr/2,t+\delta t;\vr'-\delta\vr/2,t-\delta t)$ are defined as $G^A(\vr'+\delta\vr/2,t+\delta t;\vr'-\delta\vr/2,t-\delta t)\equiv i\theta(t_2-t_1)\la [S^+_{\vr_1}(t_1),S^-_{\vr_2}(t_2)]\ra$ and 
$G^<(\vr'+\delta\vr/2,t+\delta t;\vr'-\delta\vr/2,t-\delta t)\equiv-i\la S^+_{\vr_1}(t_1)S^-_{\vr_2}(t_2)\ra$, respectively. Here, $\vr$ is defined in FI, while $\vr'$ is defined in PM. 

Let us consider the steady state in terms of time and spatially uniform interface, where $\chi^{R(<)}_{\vq\vr,\omega t}\to\chi^{R(<)}_{\vq\omega}$ and $G^{A(<)}_{\vk\vr',\omega t}\to G^{A(<)}_{\vk\omega}$. By substituting the Kadanoff Baym ansatz~\cite{Haug-text} $\xl=2i\IM\xr\fNM$ and $\Gl=2i\IM\Gr\fFM$ into Eq.~(\ref{Eq:Spin-current-01}), with $\fNM=f(\omega/T_{\rm P})$ and $\fFM=f(\omega/T_{\rm F})$ being the Bose-Einstein distribution functions in PM and FI, respectively, we obtain the general expression of spin current as follows:
\begin{eqnarray}
	I^{\rm S}
	&=& 4\Jint \!\! \ikqom \!\!\!\!\! \IM\xr\IM\Gr(\fNM-\fFM).
\label{Eq:Spin-current-02}
\end{eqnarray}
Equation (\ref{Eq:Spin-current-02}) is a spin-current version of the Meir-Wingreen formula~\cite{Meir92}, where $\Jint$ corresponds to the tunneling probability of the spin current at the interface. 
The integration of $\IM\xr$ over $\vq$ and that of $\IM\Gr$ over $\vk$ represent the density of states of the transverse spin fluctuations in PM and FI, respectively.
The difference $\fNM - \fFM$ plays a crucial role in spin-current generation and has a non-vanishing value only when the system is out of equilibrium. In the following, we investigate the difference caused by temperature difference and that by spin accumulation. 
\paragraph{Spin current driven by spin Seebeck effect.---}
First, let us consider the spin Seebeck effect~\cite{Uchida08,Uchida10,Jaworski10,Xiao10,Adachi11} that spin current injection is driven by the temperature difference $\delta T$ between PM and FI, given as $\delta T=T_{\rm P}-T_{\rm F}$. 
The difference between $\fNM$ and $\fFM$ is given by
\begin{eqnarray}
	\fNM-\fFM
    &=& \frac{\del f}{\del T}\delta T.
\label{Eq:f-difference-T}
\end{eqnarray}
Substituting Eq.~(\ref{Eq:f-difference-T}) into (\ref{Eq:Spin-current-02}), we obtain the spin-current injection due to the spin Seebeck effect as follows~\cite{Adachi11}:
\begin{eqnarray}
	I^{\rm S}
	&=& 4\Jint \!\! \ikqom \!\!\!\!\! \IM\xr\IM\Gr\frac{\del f}{\del T} \delta T.
\label{Eq:spinC-SSE}
\end{eqnarray}
%
\paragraph{Spin current driven by spin accumulation.---}
Now, let us focus on the spin-current injection driven by the spin accumulation. The expression of spin accumulation at the interface is given by $\dm=2e\alpha_{\rm SH}\rho_{\rm N}\lambda_{\rm N}j_{\rm c}\tanh(d_{\rm N}/2\lambda_{\rm N})$~\cite{Zhang00,Maekawa-text}, where $\alpha_{\rm SH}$, $\rho_{\rm N}$, $\lambda_{\rm N}$, $j_{\rm c}$, and $d_{\rm N}$ are the spin Hall angle, electrical resistivity, spin diffusion length, charge current, and thickness of metal, respectively. 
The retarded and the lesser components of the spin susceptibility in the metal, $\xr$ and $\xl$, are modified by the spin accumulation $\dm$ as $\xr \!\! \to \!\! \chi^R_{\vq,\omega+\dm}$ and $\xl \!\! \to \!\! \chi^<_{\vq,\omega+\dm}=2i\IM\chi^R_{\vq\omega} f^{\rm P}_{\omega+\dm}$, respectively. 

The difference between $f^{\rm P}_{\omega+\dm}$ and $\fFM$ is as follows:
\begin{eqnarray}
	f^{\rm P}_{\omega+\dm}-\fFM
    &=& \frac{\del f}{\del\omega}\frac{\dm}{\hbar}.
\label{Eq:f-difference-mu}
\end{eqnarray}
Substituting Eq.~(\ref{Eq:f-difference-mu}) into (\ref{Eq:Spin-current-02}), we obtain the spin-current injection driven by spin accumulation as follows:
\begin{eqnarray}
	I^{\rm S}
	&=& 4\Jint \!\! \ikqom \!\!\!\!\! \IM\xr\IM\Gr\frac{\del f}{\del \omega} \frac{\dm}{\hbar}.
\label{Eq:spinC-injection}
\end{eqnarray}
Note that Eq.~(\ref{Eq:spinC-injection}) reduces to (S10) in Ref.~\onlinecite{Kajiwara10} when we evaluate spin susceptibility in the metal $\xr$ for the noninteracting electrons.
\paragraph{Heat transport mediated by spin current.---}
Following Ref.~\onlinecite{Maki65}, we define the heat current $I^{\rm Q}$ injected into the ferromagnet as the time derivative of the Hamiltonian of the ferromagnet $H_{\rm m}$, $I^{\rm Q}\equiv\sum_{i\in F}\la \del_t H_{\rm m} \ra$, where $\la \cdots \ra$ denotes the statistical average.
The Heisenberg equation of motion for $H_{\rm m}$ gives $\del_t H_{\rm m}=(i\hbar)^{-1}[H_{\rm m},H_{\rm sd}]$. 
Substituting Eq.~(\ref{Eq:model-sd}) into $\del_t H_{\rm m}$ and taking the statistical average give the following heat current:
\begin{eqnarray}
	I^{\rm Q}
	&=& \frac{\Jsd}{2}\sum_{i\in {\rm int}}\del_{t'} \la\vsigma_i(t)\cdot\vS_i(t')\ra_{t'\to t},
\label{Eq:Heisenberg-02}
\end{eqnarray}
where we use the Heisenberg equation of motion for localized spin at the interface $\del_t\vS_i=(i\hbar)^{-1}[\vS_i,H_{\rm m}]$ to derive Eq.~(\ref{Eq:Heisenberg-02}). 

Now we consider the spin wave approximation in the lowest order of $1/S_0$ expansion, with $S_0$ being the size of the localized spins. 
The time derivative of $S^z_i$ vanishes because the z-component of the localized spins $S^z_i$ becomes constant.
By performing the perturbative calculation up to the second order of the interfacial interaction $\Jsd$, we obtain the heat current as
\begin{eqnarray}
	I^{\rm Q}
    &=& 2\Jint\RE \!\! \ikqom \!\!\!\!\! \hbar\omega[\xr\Gl+\xl\Ga].
\label{Eq:HeatC-02}
\end{eqnarray}

By substituting the Kadanoff Baym ansatz into Eq.~(\ref{Eq:HeatC-02}), we can rewrite the heat current as
\begin{eqnarray}
	I^{\rm Q}
	&=& 2\Jint\RE \!\! \ikqom \!\!\!\!\! \hbar\omega\IM\xr\IM\Gr(\fNM-\fFM).
\label{Eq:HeatC-03}
\end{eqnarray}
Especially, 
substituting Eqs.~(\ref{Eq:f-difference-T}) and (\ref{Eq:f-difference-mu}) into (\ref{Eq:HeatC-03}), we obtain the interfacial heat current caused by the temperature difference:
\begin{eqnarray}
I^{\rm Q}
	&=& 4\Jint \!\! \ikqom \!\!\!\!\! \hbar\omega\IM\xr\IM\Gr\frac{\del f}{\del T} \delta T,
\label{Eq:HeatC-Fourier} 
\end{eqnarray}
and that caused by the spin accumulation:
\begin{eqnarray}
	I^{\rm Q}
	&=& 4\Jint \!\! \ikqom \!\!\!\!\! \hbar\omega\IM\chi^R_{\vq,\omega+\dm}\IM\Gr\frac{\del f}{\del\omega} \frac{\dm}{\hbar}.
\label{Eq:HeatC-SPE}
\end{eqnarray}

Equations (\ref{Eq:spinC-SSE}), (\ref{Eq:spinC-injection}), (\ref{Eq:HeatC-Fourier}), and (\ref{Eq:HeatC-SPE}) are summarized by Onsager's reciprocal relation~\cite{DeGroot-text}:
\begin{eqnarray}
	\begin{pmatrix}
		I^{\rm S} \\
        I^{\rm Q}
	\end{pmatrix}
    =
    \begin{pmatrix}
    	L_{11} &L_{12} \\
        L_{21} &L_{22} \\
    \end{pmatrix}
    \begin{pmatrix}
		\dm \\
        \delta T/T
	\end{pmatrix},
\label{Eq:Onsager-MT}
\end{eqnarray}
where the transport coefficients are given by
\begin{eqnarray}
	L_{11}
    &=& 4\Jint\ikqom\IM\xr\IM\Gr\frac{\del f}{\del\omega}, \label{Eq:Ons-L11}\\
    L_{12}
    &=& 4\Jint\ikqom\IM\xr\IM\Gr\frac{\del f}{\del T}, \label{Eq:Ons-L12}\\
    L_{21}
    &=& 4\Jint\ikqom\hbar\omega\IM\xr\IM\Gr\frac{\del f}{\del\omega}, \label{Eq:Ons-L21}\\
    L_{22}
    &=& 4\Jint\ikqom\hbar\omega\IM\xr\IM\Gr\frac{\del f}{\del T}.\label{Eq:Ons-L22}
\end{eqnarray}
Substituting the relation $\omega\del f/\del\omega = -T\del f/\del T$ into Eq.~(\ref{Eq:Ons-L21}) yields the relation $L_{12}=L_{21}$. 
\paragraph{Temperature change at the interface.---}
Finally, we estimate the temperature change $\Delta T$ due to the spin Peltier effect. 
At the interface, magnons are excited and accumulated by the spin Peltier effect. The energy change $\Delta E$ at the interface is generated by the accumulation of magnons. Then, the temperature change $\Delta T$ is obtained as $\Delta T=\Delta E/C_{\rm FI}$, with $C_{\rm FI}$ being the heat capacity of the ferromagnet. 

Now, we formulate the energy change $\Delta E$ of the magnons with the lesser component of transverse spin susceptibility $\Gl$. 
In the spin wave approximation, the operators of localized spins are given by $S^+_i\approx\sqrt{2S_0}a_i$, $S^-_i\approx\sqrt{2S_0}a^{\dag}_i$, where $a_i$ and $a^{\dag}_i$ are the creation and the annihilation operators of the magnons. Substituting these relations into the lesser component of transverse spin susceptibility in FI, we obtain
$G^<_{\vk}(t_1,t_2)=-2i S_0\la a^{\dag}_{\vk}(t_1)a_{\vk}(t_2)\ra$. 
Because the statistical average of $a^{\dag}_{\vk}(t_1)a_{\vk}(t_2)$ can be interpreted as the number of the magnons when $t_2$ corresponds to $t_1$, the energy change $\Delta E$ is given by
\begin{eqnarray}
	\Delta E
    &\equiv&\frac{-1}{2S_0}\int_{\vk\omega} \hbar\omega \IM(G^{<}_{\vk\omega}-G^{0<}_{\vk\omega}),
\label{Eq:E-magnon}
\end{eqnarray}
where $G^{0<}_{\vk\omega}=2i\IM G^R_{\vk\omega}\fFM$ is the lesser Green's function of the free magnons. 

Let us consider a bilayer system composed of the platinum (Pt) and the yittrium iron garnet (YIG).
In spin wave approximation, the retarded component of transverse spin susceptibility $\Gr$ is given by
$\Gr=2S_0(\omega-\omega_{\vk}+i\alpha\omega)^{-1}$, where $\omega_{\vk}=A\vk^2+\gamma H_0$ is the dispersion relation of magnons, with $A$, $\gamma$, and $H_0$ being the stiffness constant, gyromagnetic ratio, and static magnetic field in YIG, respectively. $\alpha$ is the Gilbert damping constant of the magnons. 
After perturbative calculation up to the second order of $\Jsd$, we obtain the lesser Green's function of the magnons at the interface $G^{<}_{\vk\omega}$ as follows:
\begin{eqnarray}
	\Gl
    &=& G^{0<}_{\vk\omega} - 4i\frac{\Jint S_0}{\alpha\omega} \!\! \int_{\vq}\IM\xr\IM\Gr\frac{\del f}{\del \omega}\frac{\dm}{\hbar}.
\label{Eq:lesser-int}
\end{eqnarray}

Equation (\ref{Eq:lesser-int}) shows the accumulation of magnons driven by spin-current injection. Let us consider the rate equation of the magnons at the interface. Since the number density of the excited magnons can be derived from the lesser component of the transverse spin susceptibility in FI, the rate equation of magnons is written as $\del G^{<}_{\vk\omega,t}/\del t = (G^{<}_{\vk\omega,t}-G^{0<}_{\vk\omega})/\tau_{\vk\omega} - I^{\rm S}_{\vk\omega}$, with $\tau_{\vk\omega}=(\alpha\omega)^{-1}$ being the lifetime of the magnons. Here, the source term $I^{\rm S}_{\vk\omega}$ is the spin current of a particular magnon with the wavenumber $\vk$ and frequency $\omega$, defined as $I^{\rm S}_{\vk\omega}\equiv 4\Jint\int_{\vq}\IM\xr\IM\Gr(\del f/\del\omega)(\dm/\hbar)$. In the steady state, where $G^{<}_{\vk\omega,t}\to G^{<}_{\vk\omega}$ and the l.h.s of the rate equation vanishes, the rate equation reduces to $\Gl-G^{0<}_{\vk\omega}=I^{\rm S}_{\vk\omega}\tau_{\vk\omega}$, corresponding to Eq.~(\ref{Eq:lesser-int}). 

Substituting Eq.~(\ref{Eq:lesser-int}) into (\ref{Eq:E-magnon}), we obtain the energy change of the magnons as 
\begin{eqnarray}
	\Delta E
    &=& \frac{\hbar}{2\alpha}I^{\rm S},
\label{Eq:E-spinC}
\end{eqnarray}
where $I^{\rm S}$ is shown in Eq.~(\ref{Eq:spinC-injection}). 

The spin susceptibility in Pt, $\xr$, is written as $\xr=\chi_{\rm N}(\tau^{-1}_{\rm sf} + D_{\rm N}\vq^2+i\omega)^{-1}$~\cite{Adachi11}, where $\tau_{\rm sf}$ and $D_{\rm N}$ are the spin-flip time and the diffusion constant of Pt, respectively. 
By integrating $I^{\rm S}$ over $\omega$ in Eq.~(\ref{Eq:E-spinC}) by using the relation $\IM\Gr\approx-\pi\delta(\omega-\omega_{\vk})$, we have $\Delta E=-(N_{\rm int}g_{\rm s}/2)(\kB T/\hbar\omega_{\rm M})^{3/2} (\gamma_1/\gamma_2)\dm$, where $g_{\rm s}$ is given as $g_{\rm s}=(\Jsd/\hbar)^2 S_0 \int_{\vq}\IM\chi^R_{\vq,\gamma H_0}/(\gamma H_0)$, with $\omega_{\rm M}$ being the maximum energy of the magnons estimated from the Curie temperature $T_{\rm C}$ as $\omega_{\rm M}\equiv \kB T_{\rm C}/\hbar$. 
The numerical factors $\gamma_1$ and $\gamma_2$ are defined by $\gamma_1=\int^1_0 dx\int^{y_{\rm M}}_{y_0} dy y\sqrt{x(y-y_0)}[4((1+x)^2+(y\kB T\tau_{\rm sf}/\hbar)^2)\sinh^2(y/2)]^{-1}$ and $\gamma_2=\int^1_0 dx\sqrt{x}[(1+x)^2+(\gamma H_0\tau_{\rm sf})^2]^{-1}$, respectively. In the factor $\gamma_2$, $y_0$ and $y_{\rm M}$ are given by $y_0=\hbar\gamma H_0/\kB T$ and $y_{\rm M}=\hbar\omega_{\rm M}/\kB T$, respectively. 

We examine the experiment in Ref.~\onlinecite{Daimon16}. 
By using the parameters of Pt in Ref.~\onlinecite{Daimon16} as $\rho_{\rm N}=0.48$~$\mu\Omega$ $\cdot$m, $\lambda_{\rm N}=7.3$~nm~\cite{Wang14}, $j_{\rm c}=1.0\times10^9$~A/m$^2$, $d_{\rm N}=5$~nm, and $\alpha_{\rm SH}=0.013$~\cite{Sinova16}, we obtain the spin accumulation at the interface as $\dm=2.3\times10^{-8}$~eV. 
In the case of YIG, where $T_{\rm C}=565$~K and $H_0=200$~Oe, we estimate $\gamma_1=0.215$ and $\gamma_2=0.285$ at room temperature. Combining the values of $\dm$, $\gamma_1$ and $\gamma_2$, and $\alpha=10^{-5}$ and $g_{\rm s}=0.1$~\cite{Wang14}, we obtain the energy change normalized per site of localized spin at the interface $\Delta E/N_{\rm int}$ as $\Delta E/N_{\rm int}=-3.3\times10^{-5}$~eV. 
Taking $N_{\rm int}=1.0\times 10^{11}$ and $C_{\rm FI}=\tilde{c}_{\rm FI}\rho_{\rm FI}a^3_{\rm FI}N_{\rm int}$, with the density $\rho_{\rm FI}=5170$~(kg/m$^3$), the lattice constant $a_{\rm FI}=1.24\times10^{-9}$~m, and the specific heat $\tilde{c}_{\rm FI}=570$~J/(kg$\cdot$ K) of YIG, the temperature change is estimated to be $\Delta T=-1$~mK, which is consistent with the experimental result~\cite{Daimon16}. 
\paragraph{Conclusion.---}
In this study, a microscopic theory of the spin Peltier effect in a magnetic bilayer structure system consisting of PM and FI was formulated using the nonequilibrium Green's function method. 
We derived the spin- and heat-currents driven by temperature gradient as well as by spin accumulation at the interface in terms of spin susceptibility and the magnons' Green's function. 
These currents have been summarized using Onsager's reciprocal relation. 
In addition, we estimated heat generation and absorption at the interface due to spin injection from PM into FI. 
Our theory will provide a microscopic understanding of the conversion phenomena between spin and heat at the magnetic interface. 

\paragraph{Acknowledgement.---}
We are grateful to S. Daimon, M. Sato and E. Saitoh for valuable discussions. This work was financially supported by the ERATO, JST, the Grants-in-Aid for Scientific Research (Grant Nos. 26103006, 26247063, 16H04023, 15K05153) from JSPS and the MEXT of Japan.



\begin{thebibliography}{99}
%
\bibitem{Uchida08}
	K. Uchida, S. Takahashi, K. Harii, J. Ieda, W. Koshibae, K. Ando, S. Maekawa, and E. Saitoh, 
	Nature (London) {\bf 455}, 778 (2008). 

\bibitem{Jaworski10}
	C. M. Jaworski, J. Yang, S. Mack, D. D. Awschalom, J. P. Heremans, and R. C. Myers, 
	Nature Mater. {\bf 9}, 898 (2010). 
    
\bibitem{Uchida10}
	K. Uchida, J. Xiao, H. Adachi, J. Ohe, S. Takahashi, J. Ieda, T. Ota, Y. Kajiwara, H. Umezawa, H. Kawai, G. E. W. Bauer, S. Maekawa, and E. Saitoh, 
	Nature Mater. {\bf 9}, 894 (2010). 

\bibitem{Xiao10}
	J. Xiao, G. E. W. Bauer, K. C. Uchida, E. Saitoh, and S. Maekawa, 
	Phys. Rev B, {\bf 81}, 214418 (2010). 

\bibitem{Adachi11}
	H. Adachi, J. I. Ohe, S. Takahashi, and S. Maekawa, 
    Phys. Rev B, {\bf 83}, 094410 (2011).

\bibitem{Flipse14}
	J. Flipse, F. K. Dejene, D. Wagenaar, G. E. W. Bauer, J. B. Youssef, and B. J. van Wees, 
	Phys. Rev. Lett., {\bf 113}, 027601 (2014). 
	
\bibitem{Daimon16}
	S. Daimon, R. Iguchi, T. Hioki, E. Saitoh, and K. Uchida, 
	Nature Commun. {\bf 7}, 13754 (2016). 

\bibitem{Gravier06}
	L. Gravier, S. Serrano-Guisan, F. Reuse, and J. P. Ansermet,  
	Phys. Rev. B, {\bf 73}, 024419 (2006). 

\bibitem{Hatami09}
	M. Hatami, G. E. W. Bauer, Q. Zhang, and P. J. Kelly, 
	Phys. Rev. B, {\bf 79}, 174426 (2009). 
    
\bibitem{Kovalev09}
	A. A. Kovalev and Y. Tserkovnyak 
	Phys. Rev. B, {\bf 80}, 100408(R) (2009). 

\bibitem{Bauer10}
	G. E. W. Bauer, S. Bretzel, A. Brataas, and Y. Tserkovnyak, 
	Phys. Rev. B, {\bf 81}, 024427 (2010). 
    
\bibitem{Basso16}
	V. Basso, E. Ferraro, A. Magni, A. Sola, M. Kuepferling, and M. Pasquale, 
	Phys. Rev. B, {\bf 93}, 184421 (2016). 

\bibitem{Sinova16}
	J. Sinova, S. O. Valenzuela, J. Wunderlich, C. H. Back, and T. Jungwirth, 
	Rev. Mod. Phys., {\bf 87}, 1213 (2015). 

\bibitem{Konig03}
	J. K\"onig and J. Martinek,
	Phys. Rev. Lett., {\bf 90}, 166602 (2003). 

\bibitem{Meir92}
	Y. Meir and N. S. Wingreen,
	Phys. Rev. Lett., {\bf 68}, 2512 (1992). 

\bibitem{Haug-text}
	H. Haug and A. -P. Jauho,
	{\it Quantum Kinetics in Transport and Optics of Semiconductors}
	(Springer-Verlag, Berlin, 1996). 

\bibitem{Zhang00}
	S. Zhang,
	Phys. Rev. Lett., {\bf 85}, 393 (2000).

\bibitem{Maekawa-text}
	 S. Maekawa, S. O. Valenzuela, E. Saitoh, and T. Kimura,
	{\it Spin Current}
	(Oxford University Press, Oxford, 2012).

\bibitem{Kajiwara10}
	Y. Kajiwara, K. Harii, S. Takahashi, J. Ohe, K. Uchida, M. Mizuguchi, H. Umezawa, H. Kawai, K. Ando, K. Takanashi, S. Maekawa, and E. Saitoh
	Nature (London) {\bf 464}, 262 (2010).

\bibitem{Maki65}
	K. Maki and A. Griffin, 
    Phys. Rev. Lett., {\bf 15}, 921 (1965). 
    
\bibitem{DeGroot-text}
	S. R. de Groot and P. Mazur, 
    {\it Non-Equilibrium Thermodynamics}
	(Dover, New York, 1984). 

\bibitem{Wang14}
	H. L. Wang, C. H. Du, Y. Pu, R. Adur, P. C. Hammel, and F. Y. Yang,
	Phys. Rev. Lett., {\bf 112}, 197201 (2014).
\end{thebibliography}
\end{document}